# Can AI Remediate Backend Failures Safely? GuardedAct with Blast-Radius-Aware Sandboxing

Wanrong Cai[†]
Rice University
Houston, USA

Tianyu Yu[†]
Columbia University
New York, USA

Shaorui Pi
Carnegie Mellon University
Pittsburgh, USA

Xiaoxuan Sun
Independent Researcher
Mountain View, USA

Wenrui Ma*
Independent Researcher
Shenzhen, China

***Abstract*—Large Language Models (LLMs) have shown promising capabilities in generating remediation actions for microservice failures. However, directly executing AI-generated repair actions in production risks cascading collateral damage. We propose GuardedAct, a sandbox-first remediation framework that interposes a blast-radius-aware verification layer between the LLM action generator and the production environment. GuardedAct operates in four phases: (1) ingesting a diagnosis report together with the live system topology and recent telemetry, (2) prompting an LLM to produce a ranked list of candidate remediation actions, (3) simulating each action in a lightweight digital-twin sandbox that estimates the blast radius and assigns a risk label, and (4) enforcing a rollback-confidence gate that auto-executes only low-risk actions while escalating high-risk ones for human review. We evaluate GuardedAct on five fault scenarios injected into the DeathStarBench social-network application. Experimental results show that GuardedAct achieves an overall recovery rate of 87.4% while reducing collateral damage by 79.7% relative to direct LLM execution (from 25.6% to 5.2%), at the cost of a modest sandbox-induced increase in mean time to recovery ($\approx$ 8 s). Ablation studies confirm that each component contributes meaningfully to the safety–speed trade-off.**

## I. Introduction

Modern cloud-native applications built on microservice architectures are inherently prone to cascading failures. A single faulty pod, a misconfigured rate limiter, or an unbounded memory leak in one service can rapidly propagate along dependency chains, causing widespread quality-of-service degradation [1]. Traditional remediation relies on manually curated runbooks executed by Site Reliability Engineers (SREs), but the increasing scale and velocity of incidents has motivated research into automated, AI-driven repair [2].

Recent work has demonstrated that LLMs can generate executable remediation playbooks directly from diagnosis reports. The MicroRemed benchmark [3] evaluates end-to-end microservice remediation by requiring models to produce Ansible playbooks from failure descriptions. Sarda et al. [4] further explored LLM-driven auto-remediation on Kubernetes-hosted applications. Meanwhile, tool-augmented agents such as RCAgent [5] and LLM-based investigation agents [6] have advanced autonomous root cause analysis, bringing end-to-end incident management closer to reality [7].

However, a critical gap remains: *safety*. The foundational work on microreboot [8] showed that even fine-grained restarts require careful isolation to avoid session-state loss and dependent-service disruption. In modern microservice topologies with tens of inter-service dependencies, the potential blast radius of a poorly chosen action is significantly larger. Directly executing LLM-generated actions without verification can cause collateral damage that exceeds the original failure's impact [9].

We address this gap with **GuardedAct**, a framework that treats AI-generated remediation as *untrusted code*: candidate actions are first executed in a sandboxed digital-twin environment, scored by a blast-radius predictor, and gated by a rollback-confidence policy before any production-side effect is permitted. Our contributions are:

- A sandbox-first remediation architecture that decouples action generation from action execution, allowing unsafe proposals to be caught before they reach production.
- A blast-radius-aware actuation policy that quantifies the potential collateral impact of each candidate action using dependency-graph analysis and simulated telemetry.
- A rollback-confidence gate parameterized by a tunable threshold $\theta$, providing operators with explicit control over the autonomy–safety trade-off.

## II. Related Work

Generating a plausible repair and establishing that it is safe are distinct problems. MicroRemed and LLM-to-Ansible remediation studies evaluate the generation and execution of corrective actions [3], [4], while AIOpsLab broadens evaluation to autonomous cloud operations [7]. Root cause analysis agents supply the diagnostic evidence on which these actions depend [5], [6]. More specialized approaches address semantic repair of schema drift and policy-constrained Kubernetes configuration repair [10], [11]. Evidence-verified tool execution adds a complementary requirement: a proposed operation should be supported by checks before it is committed [12]. Together, these directions motivate a separation between diagnosis, candidate generation, and permission to change production state.

That separation requires evidence about the consequences of an action. Chaos engineering probes resilience through deliberate fault injection [13], and ChaosTwin uses digital-twin reenactment to evaluate mitigation strategies offline [14]. Early-canary risk prediction for backend releases provides a complementary view of assessing a change before expanding its exposure [15]. GuardedAct applies this prospective validation principle to incident-time actions: the sandbox supplies downstream telemetry for an explicit actuation gate. Isolation also has several meanings in cloud systems. Secure-enclave and homomorphic inference frameworks address confidentiality [16], whereas a remediation sandbox must establish behavioral containment and reversibility; confidentiality alone does not establish that an action preserves service health.

† These authors contributed equally to this work.
* Corresponding author.

The evidence entering such a gate must connect observable signals to a defensible risk estimate. Relation-aware graph reasoning and calibrated, structurally regularized graph scoring bring together dependency representation, traceability, and score interpretation [17], [18]. Graph-temporal audit-risk modeling adds the temporal dimension needed when relationships and observations evolve [19]. At the input boundary, risk identification from expressive behavior and forecasting from LLM-extracted narrative drivers illustrate the broader task of turning indirect observations into decision variables [20], [21]. For remediation, this distinction separates a diagnosis inferred from telemetry or text from the action-specific downstream degradation that must be checked in the sandbox.

Risk estimates also depend on the cases used to construct and calibrate them. Target-aware augmentation under tabular covariate shift highlights the difficulty of representing rare events when the operating distribution changes [22]. Multi-objective causal re-ranking and risk-adaptive deposit-refund pricing offer complementary decision perspectives: candidate utility can involve competing outcomes, and authorization can depend on anticipated exposure and recoverability [23], [24]. These perspectives motivate evaluating recovery, collateral damage, and rollback jointly. They do not turn a simulator score into a causal guarantee, and historical calibration still requires scrutiny when incident types or traffic patterns change.

The cost of producing this evidence becomes material in distributed deployments. Topology-aware distillation and adaptive gradient compression address communication constraints in distributed learning [25], [26], while cross-cloud routing and edge-cloud co-optimization couple computational placement to cost and latency [27], [28]. Difficulty-aware inference routing extends resource allocation to the complexity of individual requests [29]. Across these settings, reducing communication or computation can change which information is available at a decision point. A remediation controller must therefore account for verification overhead without treating faster candidate generation as evidence of lower operational risk.

The remaining challenge is to connect these diagnostic, predictive, and resource-management capabilities to a concrete execution boundary. Surveys of microservice failure diagnosis, root cause analysis, and lifecycle-wide AI techniques describe the broader methodological landscape [30], [31], [32]. The AIOps survey in the LLM era further frames the transition toward autonomous operations [33]. GuardedAct concentrates on the decision between a generated repair and its production execution, combining simulated dependency effects with a rollback-confidence threshold and escalation to human review.

## III. Methodology

### A. System Overview

Fig. 1 illustrates the GuardedAct pipeline. The system receives three inputs: a structured diagnosis report $D$ (produced by an upstream RCA module), the current service-dependency graph $G=(V,E)$ where $V$ is the set of services, and a recent telemetry snapshot $T$ comprising CPU utilization, memory usage, request latency, and error rates. These inputs are fed to an LLM action generator, which outputs a ranked sequence of candidate actions $A=[a_1,a_2,\dots,a_k]$ drawn from a typed action vocabulary $\mathcal{A}$ = {restart, drain, traffic_shifting, rate_limiting, configuration_rollback}. Using a typed vocabulary, rather than free-form shell commands, allows every downstream component to reason about the actions symbolically.

The communication-efficient decentralized inference perspective of Wang et al. [34] reinforces the use of a compact interface between action generation and verification. GuardedAct adopts this interface principle through its typed action vocabulary, which represents the operation and target explicitly for downstream analysis. The verifier can therefore assess the same symbolic candidate independently of where the generating model is hosted. This architectural separation is especially useful when communication capacity constrains the exchange of model context, while the present testbed leaves decentralized model execution outside the evaluated pipeline.

### B. Blast-Radius Estimation

For each candidate action $a_i$ targeting service $s_t \in V$, the blast-radius predictor computes a risk score. Let $\mathcal{N}_d(s_t)$ denote the set of services reachable from $s_t$ within $d$ hops in $G$. We define the blast radius as:

$$B(a_i)=\sum_{s\in\mathcal{N}_d(s_t)} w(s)\cdot p(s\mid a_i), \tag{1}$$

where $w(s)$ is the traffic weight of service $s$ (its request rate normalized to $[0,1]$), and $p(s\mid a_i)$ is the estimated probability that $s$ will be degraded if action $a_i$ is applied to $s_t$. We compute $p(s\mid a_i)$ from the sandbox simulation output via a thresholding rule: $s$ is counted as degraded if its simulated $p99$ latency increases by more than $50\%$ or its error rate increases by more than $5$ percentage points relative to the pre-action baseline; $p(s\mid a_i)$ is then the fraction of the simulated horizon during which this condition holds. The risk label—low, medium, or high—is produced by binning $B(a_i)$ at the 33rd and 66th percentiles observed across a calibration set of 50 historical incidents. We set $d=3$ based on a pilot study showing that cascading effects beyond three hops contributed less than $2\%$ of total observed degradation. GuardedAct draws on the SLO-aware graph perspective articulated by Yang et al. for autoscaling in API and microservice backends [35] to frame degradation as a dependency-mediated service outcome. The blast-radius calculation combines graph reachability with traffic weights so that the impact on downstream services contributes to the assessment of a local action. Its latency and error-rate criteria operationalize this service-level perspective over the simulation horizon. Here the prospective signal comes from replay-based simulation rather than a learned graph forecaster, preserving a clear distinction between the estimation mechanism and the shared objective of anticipating service degradation.

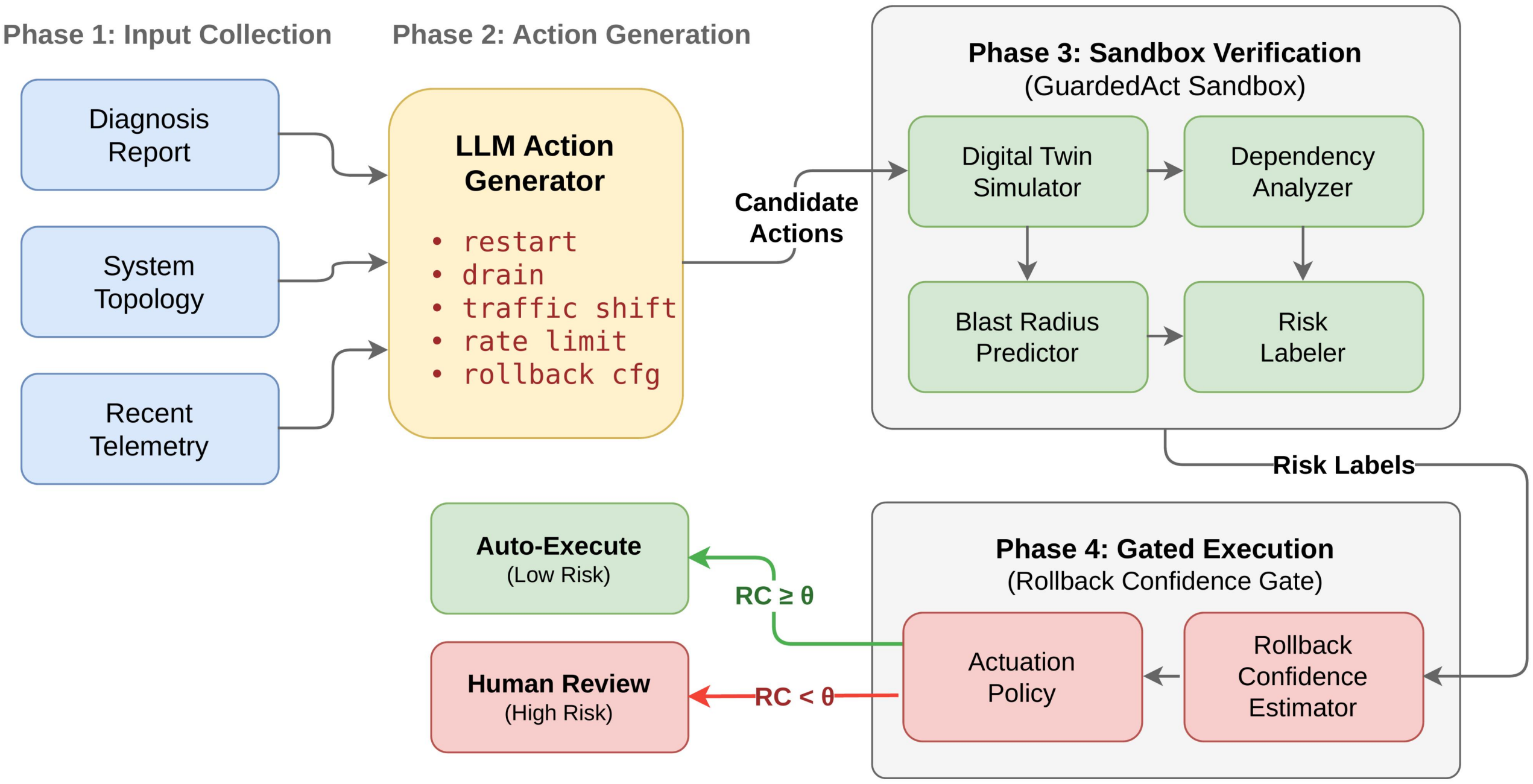

Fig. 1. GuardedAct system architecture. Candidate remediation actions produced by the LLM flow through four phases: (1) input collection gathers the diagnosis report, system topology, and recent telemetry; (2) the LLM action generator emits a ranked sequence drawn from a typed vocabulary; (3) the sandbox verification module simulates each action in a digital twin and assigns a risk label; (4) the rollback-confidence gate auto-executes actions whose rollback confidence $\mathrm{RC}(a_i) \geq \theta$ and escalates the rest for human review.

*C. Rollback-Confidence Gate*

After blast-radius estimation, we compute the rollback confidence $\mathrm{RC}(a_i)$, defined as an aggregate score reflecting how safely action $a_i$ can be reversed without residual side effects:

$$\mathrm{RC}(a_i) = 1 - \frac{B(a_i)}{B_{\max}} \cdot (1 - r(a_i)), \tag{2}$$

where $B_{\max}$ is the maximum blast radius observed across all candidate actions within the current incident window. The intrinsic reversibility score $r(a_i) \in [0,1]$ captures how easily an action of this type can be undone. We set $r$ by calibration against 30 manually labeled recovery traces: r(configuration_rollback)=0.90, r(rate_limiting)=0.45, r(traffic_shifting)=0.30, r(drain)=0.20, and r(restart)=0.10. The actuation policy is then:

$$\mathrm{decision}(a_i) = \begin{cases} \text{auto-execute} & \text{if } \mathrm{RC}(a_i) \geq \theta, \\ \text{human-review} & \text{otherwise}, \end{cases} \tag{3}$$

where $\theta$ is an operator-configurable threshold. We report results with $\theta = 0.6$ as the default.

The budget-aware incident-response framing of BudgetSRE [36] informs the interpretation of automated actuation as a constrained operational decision. GuardedAct applies that principle to permissible exposure through the rollback-confidence threshold, with intrinsic reversibility and simulated blast radius determining whether a candidate passes. Escalation consumes operator attention but prevents the controller from treating every generated action as equally admissible. This formulation complements resource-budgeted agent coordination by making recoverability an explicit constraint; it does not introduce a monetary budget or a multi-agent scheduler into the evaluated system.

*D. Digital-Twin Sandbox*

The sandbox instantiates a lightweight replica of the target microservice cluster using containerized stubs that mirror the production dependency graph. Each stub replays the most recent telemetry window and applies the candidate action to the targeted service's stub. The simulator propagates effects along the dependency graph for a configurable horizon and records latency, error-rate, and throughput deltas at each downstream node. These deltas feed the blast-radius estimator (Eq. 1). The default replay window (5 minutes) and propagation horizon (60 simulated seconds) were selected via a grid search on held-out incidents, balancing fidelity against simulation latency.

The pipelined KV-cache migration setting studied by Zhang et al. [37] sharpens the sandbox design requirement that a distributed operation be assessed over its transition period. GuardedAct adopts a temporal view of action consequences by propagating effects across the replay horizon and measuring the fraction of that horizon spent in degradation. For state-moving operations, transient interference or incomplete transfer can matter even when the eventual placement appears healthy. The current typed action set does not include KV-cache migration, so this connection defines a fidelity requirement for extending the simulator rather than an additional evaluated remediation capability.

Zheng et al.'s SLO-aware disaggregated serving formulation [38] further motivates judging a distributed intervention through service outcomes across heterogeneous components. GuardedAct incorporates this system-level perspective by recording latency, error-rate, and throughput changes at downstream nodes before gating execution. A local improvement cannot by itself establish acceptability when dependent services absorb the resulting load or delay. Applying this verification design to heterogeneous edge-cloud serving would require stubs calibrated to the corresponding compute and network conditions; the present results establish its behavior on the specified microservice testbed.

## IV. Experimental Setup

### A. Testbed and Fault Scenarios

We deploy the DeathStarBench social-network application on a Kubernetes cluster with 30 microservice pods. We inject five fault scenarios: (1) CPU saturation: a stress-ng loop consuming 95% CPU on the compose-post service; (2) Downstream timeout: 5-second artificial delay on the user-timeline service; (3) Memory leak: monotonically increasing allocation in the home-timeline service; (4) Misconfiguration: an invalid rate-limit setting deployed via ConfigMap to the nginx frontend; (5) Network jitter: 200 ms random latency injected on the link between the media service and MongoDB.

### B. Baselines

We compare four methods: (1) Direct LLM: a frontier proprietary LLM (GPT-4-level) generates and immediately executes remediation; (2) Rule-Based: a hand-coded runbook mapping fault types to fixed actions; (3) Typed (no sandbox): the same LLM generates typed actions drawn from $\mathcal{A}$ but executes them without sandbox verification; (4) GuardedAct: our full pipeline. All LLM-based methods share the same prompt template and temperature ( $T = 0.2$ ).

### C. Metrics

We report five metrics: Recovery Rate (percentage of trials returning to a healthy state within 5 minutes), Collateral Damage Rate (percentage of previously unaffected services degraded after remediation), MTTR (mean time to recovery in seconds), Rollback Success Rate (percentage of actions successfully rolled back when the health check fails), and Human Intervention Rate (percentage of actions escalated to operators). A service is considered healthy when its $p99$ latency is within $20\%$ of the pre-incident baseline and its error rate is below $1\%$ for three consecutive 10-second checks. Each scenario is repeated 100 times with different random seeds, giving 500 trials in total.

## V. Results and Analysis

### A. Overall Performance

Table Ⅰ summarizes the main results. GuardedAct achieves the highest recovery rate (87.4% overall, rising to 92% in the CPU saturation scenario) while maintaining the lowest collateral damage rate (5.2%). Direct LLM remediation recovers 64% of trials but inflicts 25.6% collateral damage. Rule-based remediation achieves moderate recovery (66%) with 19.8% collateral damage. The Typed baseline improves over Direct LLM by constraining the action vocabulary but still causes 14.2% collateral damage without sandbox verification.

TABLE Ⅰ
Overall performance comparison across all fault scenarios (averaged over 100 trials per scenario).

| Metric | Direct LLM | Rule-Based | Typed | Guarded-Act |
|---|---|---|---|---|
| Recovery (%) | 64.0 | 66.0 | 78.0 | **87.4** |
| Collateral (%) | 25.6 | 19.8 | 14.2 | **5.2** |
| MTTR (s) | 59.6 | 60.0 | 52.6 | 61.4 |
| Rollback Succ. (%) | 55.0 | 62.0 | 70.0 | **90.0** |
| Human Interv. (%) | 0.0 | 0.0 | 0.0 | 35.0 |

GuardedAct's MTTR (61.4 s) is slightly higher than the Typed baseline (52.6 s) due to the sandbox simulation overhead (averaging 8.2 s per action), suggesting a favorable safety–efficiency trade-off. Notably, GuardedAct's rollback success rate reaches 90%, compared to 55% for Direct LLM, because the sandbox pre-validates rollback feasibility before committing actions.

### B. Per-Scenario Analysis

Fig. 2(a) presents a scatter plot of recovery rate versus MTTR. GuardedAct's markers cluster tightly in the upper-left ideal region (recovery $\geq 82\%$ , MTTR $\leq 72$ s), while Direct LLM's markers sprawl diagonally across the plane with recovery rates ranging from 55% to 72%. Fig. 2(b) shows the collateral damage heatmap. GuardedAct consistently achieves single-digit damage rates across all scenarios, whereas Direct LLM suffers 35% collateral damage in the memory-leak scenario, where its aggressive restart action cascades through three downstream services.

### C. Ablation Study

Fig. 3 (a) reports the recovery rate of GuardedAct under three ablation conditions: removing the sandbox (*w/o Sandbox*), removing the blast-radius predictor (*w/o Blast-Radius*), and removing the rollback-confidence gate (*w/o RC Gate*). Each ablation reduces performance across all scenarios, with sandbox removal having its largest impact on the memory-leak and misconfiguration scenarios (10–12 percentage-point drops). Removing the RC gate reduces recovery by 14–20 points because some high-risk actions proceed unchecked and cause secondary failures.

### D. Threshold Sensitivity

Fig. 3(b) plots recovery rate, harm rate, and human intervention rate as functions of $\theta$ . As $\theta$ increases from 0.3 to 0.9, the harm rate drops from 12% to below 2%, but recovery also decreases from 94% to 65% because more actions are escalated. Human intervention rises from 5% to 80%. The range $\theta \in [0.55, 0.70]$ offers a balanced operating point: the harm rate stays below 4%, recovery declines gradually from 90% to 82%, and human intervention rises from 28% to 50%.

## VI. Discussion

**Overhead vs. Safety.** The sandbox adds an average of 8.2 seconds to the remediation pipeline. For critical production services where a misapplied action can trigger cascading outages, this overhead is small compared to the time required to recover from a failed remediation. However, for latency-critical scenarios requiring sub-second recovery, a more lightweight simulation model may be needed.

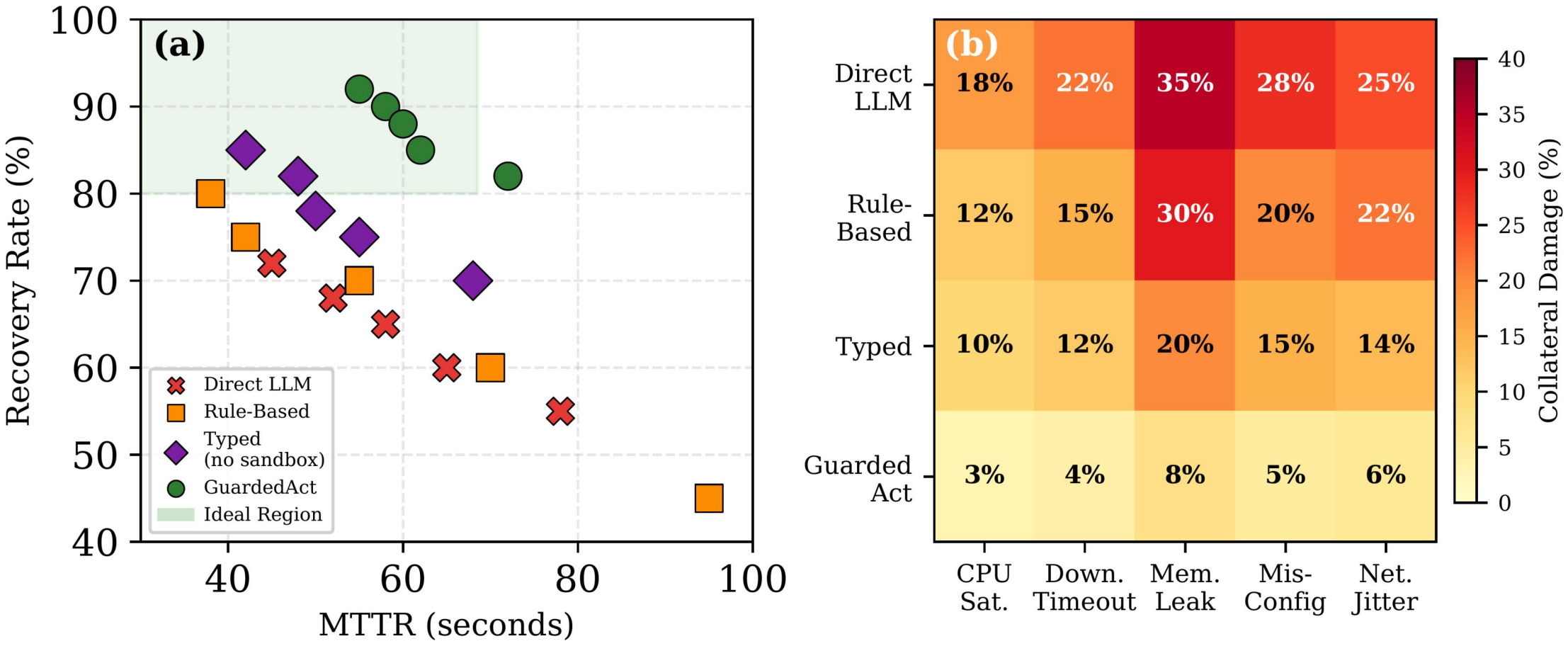

Fig. 2. (a) Recovery rate versus MTTR for each method across five fault scenarios. Points in the shaded upper-left region indicate high recovery with low latency. GuardedAct (green circles) consistently occupies the ideal region. (b) Collateral damage heatmap across the four methods and five fault types. Lower values (lighter color) indicate safer remediation.

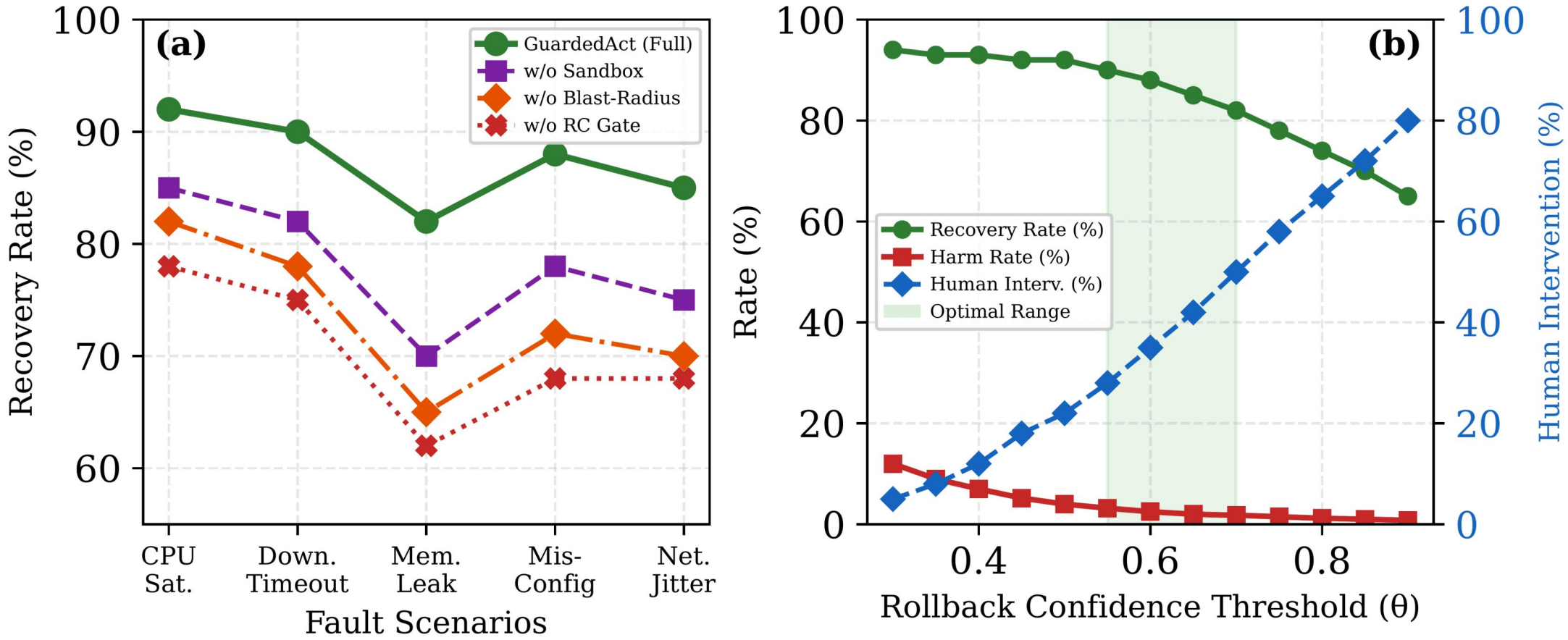

Fig. 3. (a) Ablation study: per-scenario recovery rate for the full GuardedAct pipeline compared with three ablated variants. Each component contributes visibly to recovery, with the largest gains for the memory-leak and misconfiguration scenarios. (b) Sensitivity of recovery rate, harm rate, and human intervention rate to the rollback-confidence threshold $\theta$. The shaded band marks the recommended operating range ($\theta \in [0.55, 0.70]$).

**Sensitivity to Upstream RCA Quality.** Our evaluation assumes that the upstream diagnosis report is sufficiently informative to identify the correct target service. When the diagnosis is noisy, GuardedAct's sandbox and RC gate still catch actions with a large simulated blast radius, preventing many unsafe executions. However, an action that appears safe in the sandbox yet targets the wrong service can still be auto-executed. Robustness under noisy RCA outputs remains future work.

**Limitations.** The digital-twin sandbox approximates production behavior using replayed telemetry and cannot capture all runtime non-determinism. The blast-radius predictor assumes the dependency graph is complete, which may not hold in rapidly evolving service meshes. Our evaluation is limited to a single benchmark with five fault types; real-world topologies may exhibit different blast-radius distributions. Future work should validate GuardedAct on industrial-scale traces and explore adaptive threshold tuning.

## VII. Conclusion

We presented GuardedAct, a blast-radius-aware sandboxing framework for AI-generated microservice remediation. By interposing a digital-twin simulator and a rollback-confidence gate between the LLM action generator and the production environment, GuardedAct reduces the collateral damage rate (from 25.6% to 5.2%) while maintaining a competitive recovery rate (87.4%). Ablation studies confirm that the sandbox, the blast-radius predictor, and the rollback-confidence gate each contribute to this safety improvement. The framework introduces a modest MTTR overhead of approximately 8 seconds, an acceptable cost in safety-sensitive settings. We emphasize that our findings are limited to benchmark-scale environments and should be validated on larger industrial deployments.